\documentclass[]{aastex631}

\usepackage{float}
\usepackage{array}
\usepackage{longtable}

\begin{document}

\title{Directly Characterizing Dome Seeing:\\ Differential Image Motion Sensor Using Multisources (DIMSUM)}

\author[0000-0002-4303-0512]{Ali Kurmus}
\affiliation{Department of Physics \& Department of Astronomy, Harvard University, Cambridge MA USA 02138}
\author[0000-0002-1311-4942]{Elana Urbach}
\affiliation{Department of Physics, Harvard University, Cambridge MA USA 02138}
\author[0000-0002-1311-4942]{Christopher W. Stubbs}
\affiliation{Department of Physics \& Department of Astronomy, Harvard University, Cambridge MA USA 02138}



\begin{abstract}

Image degradation impedes our ability to extract information from astronomical observations. One factor contributing to this degradation is ``dome seeing", the reduction in image quality due to variations in the index of refraction within the observatory dome. Addressing this challenge, we introduce a novel setup—DIMSUM (Differential Image Motion Sensor Using Multisources)—which offers a simple installation and provides direct characterization of local index of refraction variations. This is achieved by measuring differential image motion using strobed imaging that effectively``freezes" the atmosphere, aligning our captured images with the timescale of thermal fluctuations, thereby giving a more accurate representation of dome seeing effects. Our apparatus has been situated within the Auxiliary Telescope of the Vera C. Rubin Observatory. Early results from our setup are encouraging. Not only do we observe a correlation between the characteristic differential image motion (DIM) values and local temperature fluctuations (a leading cause of variations in index of refraction), but also hint at the potential of DIM measures to characterize dome seeing with greater precision in subsequent tests. Our preliminary findings underscore the potential of DIMSUM as a powerful tool for enhancing image quality in ground-based astronomical observations. Further refinement and data collection will likely solidify its place as a useful component for managing dome seeing in major observatories like the Vera C. Rubin Observatory.

\end{abstract}



\section{Introduction} \label{sec:intro}
 
In astronomy, ``seeing'' refers to the degradation of image quality caused by various atmospheric and environmental conditions, which prevent ground-based telescopes from reaching the theoretical diffraction limit. Dome seeing, in particular, involves localized variations in the index of refraction within the telescope enclosure, leading to image degradation. Among the diverse factors impacting overall image quality, dome seeing is a significant one \citep{Stubbs_2021, Hickson_2014, Hardy_1998}.

The motivation to identify and mitigate sources of image degradation stems from their substantial impact on the capabilities of observational systems. For instance, a 10-meter diameter telescope observing at optical wavelengths ($\lambda \approx 500\rm{nm}$) typically experiences a full width at half-maximum (FWHM) that is approximately 100 times worse than its diffraction limit in the absence of adaptive optics correction. Even after accounting for optomechanical factors, the effects of upper atmosphere and dome seeing persist. While airflow in the upper atmosphere remains beyond our control, optimizing the thermal conditions, cooling, and airflow within the dome is feasible. Additionally, while upper atmosphere seeing can be corrected using adaptive optics, not all ground-based telescopes, like the Vera C. Rubin Observatory with its wide field of view, can employ such methods. These considerations underscore the importance of  optimizing image quality by reducing dome seeing.

Previous studies have assessed local-seeing contributions within the telescope dome, primarily using indirect methods. These involved correlating physical parameters with the delivered image FWHM and comparing observations from both inside and outside the dome \citep{Woolf_1979, Racine_1991, Zago_1997, Lai_2019, Gilda_2022}. Alongside these indirect methods, direct optical determinations of turbulence within the dome were introduced by \citet{Lai_2019} and \citet{Bustos_Tokovinin_2018}.

\citet{Stubbs_2021} presents a method for characterizing dome seeing through strobed imaging. By installing this system along the light path inside the dome, it is possible to distinguish wavefront errors caused by the upper atmosphere from perturbations occurring in the final few meters of the optical path within the telescope dome. Comparing the centroid motion of the strobed artificial images taken within the dome, the FWHM of stellar images on the focal plane, and data from external systems allows for the isolation of various contributors to image degradation.

This method effectively probes scintillation and image centroid position changes due to in-dome optical path perturbations, thus providing accurate and real-time assessment of dome seeing. While there are various systems capable of characterizing dome seeing, what distinguishes this method is its simplicity and ease of setup. 

In this paper, we introduce the Differential Image Motion Sensor Using Multisources (DIMSUM), which measures the differential image motion between pairs in an array of strobed light sources. Our objective is to obtain a measurable quantity that, both temporally and spatially, consistently correlates with dome seeing. Such a system, even without the need of quantitative absolute calibration, could facilitate iterative correction of dome seeing by adjusting dome airflow. The strobed light source allows us to ``freeze'' the index distribution on timescales shorter than the correlation time $\tau$. Longer exposures would lead to suppression of centroid motion due to averaging effects. 

This paper is structured as follows. In Sec.~\ref{sec:theory} we outline the relevant theory regarding how DIMSUM can characterize dome seeing. In Sec.~\ref{sec:experimental_setups} we describe a conceptual version of our setup. In Sec.~\ref{sec:analysis} we explain the process to calculate differential image motion from the raw data. In Sec.~\ref{sec:early_results} we present early results. In Sec.~\ref{sec:discussion} we discuss the implications of our findings, the limitations of our current setup, and next steps.

\section{Theory}\label{sec:theory}
DIMSUM measures the differential image motion between an array of strobed light sources. We provide detailed descriptions of our setup and the methodology for calculating differential image motion in Sec. \ref{sec:experimental_setups} and \ref{sec:analysis}, respectively. This motion arises from variations in the refractive index gradients perpendicular to the light path, similar to the principles a Differential Image Motion Monitor (DIMM) employs.

As the separation between the light sources increases, we expect a corresponding increase in differential motion between them. This increase results from the diverging paths of ray bundles from the sources to the lens. As these paths diverge, they integrate over a wider length scale of optical turbulence, leading to greater variations in the refractive index gradient, and corresponding more differential image motion. 

However, we anticipate the differential image motion will reach a saturation point beyond a certain source separation. As the pairwise separation increases, the paths of the ray bundles diverge to such an extent that the turbulence each experiences becomes uncorrelated. We refer to this saturation-separation as the atmospheric coherence length \citep{Hickson_2014}. One of the key outputs from DIMSUM, like the differential image motion at this coherence length or a similar metric, can effectively correlate with dome seeing. This is related to the Fried parameter $r_o$. 

The standard parameterization of the structure of index perturbations in the atmosphere is to use a power law, $E(k)=Ck^{-5/3}$, with fluctuations scaling as a power law vs. spatial wavenumber $k$ \citep{Hickson_2014}. This model breaks down on small spatial scales once viscosity becomes a major factor. An ``outer scale'' is invoked to indicate where the power law has a break in slope at on large spatial scales. Dual-aperture differential image motion monitors measure the wavefront angle of arrival fluctuations at a single fixed spatial separation, averaged over the exposure time. The power law is presumed to properly characterize turbulence in the upper atmosphere, and so a measurement at a single $k$ value suffices to compute the consequent upper atmosphere contribution to the seeing for a large-aperture telescope. 

Index of refraction perturbations within the telescope enclosure, dome seeing, is subject to complicated boundary conditions, both thermal and mechanical, as well as complicated airflow through the slit. There is no reason to expect the assumptions that underlie the Kolmogorov model to hold. DIMSUM measures the index variations over a range of spatial scales and is taking an integral of $E(k)$ over a corresponding range of $k$. The differential image motion we measure is a lower bound on the dome seeing contribution to image degradation, but lack of knowledge of the $E(k)$ power spectrum prevents us from extrapolating the DIMSUM measurements to larger length scales. 

Our immediate goal with DIMSUM is to therefore measure a parameter that has consistent temporal and spatial correlation with dome seeing. To validate our approach, we compare our measurements against parameters closely related to refractive index structure constant, $C_{N}^2$, which is known to be representative of dome seeing. \citet{Zago_1997} outlines that temperature standard deviation ends up being the dominant factor affecting $C_{N}^2$ in settings akin to the Vera C. Rubin Observatory. Therefore, finding a correlation between our measurements and the temperature standard deviation would confirm the validity of our DIMSUM data. We note that while the temperature standard deviation is useful for validating our measurements, it cannot replace DIMSUM in characterizing dome seeing. DIMSUM's strategic positioning along the optical path enables it to directly measure the effects of dome seeing, unlike the temperature standard deviation, which provides only a local measure of index fluctuations at one point. 

\section{Experimental Setup}\label{sec:experimental_setups}
We attempt to build an instrument that can monitor image motion due to index of refraction perturbations along the light path within the dome using differential image motion as outlined in Sec.~\ref{sec:theory}. This section details a conceptual overview of the components of our setup and describes its physical installation.

\subsection{Conceptual Setup} \label{sec:conceptual_setup}
\begin{figure}
    \centering
    \includegraphics[width=\textwidth]{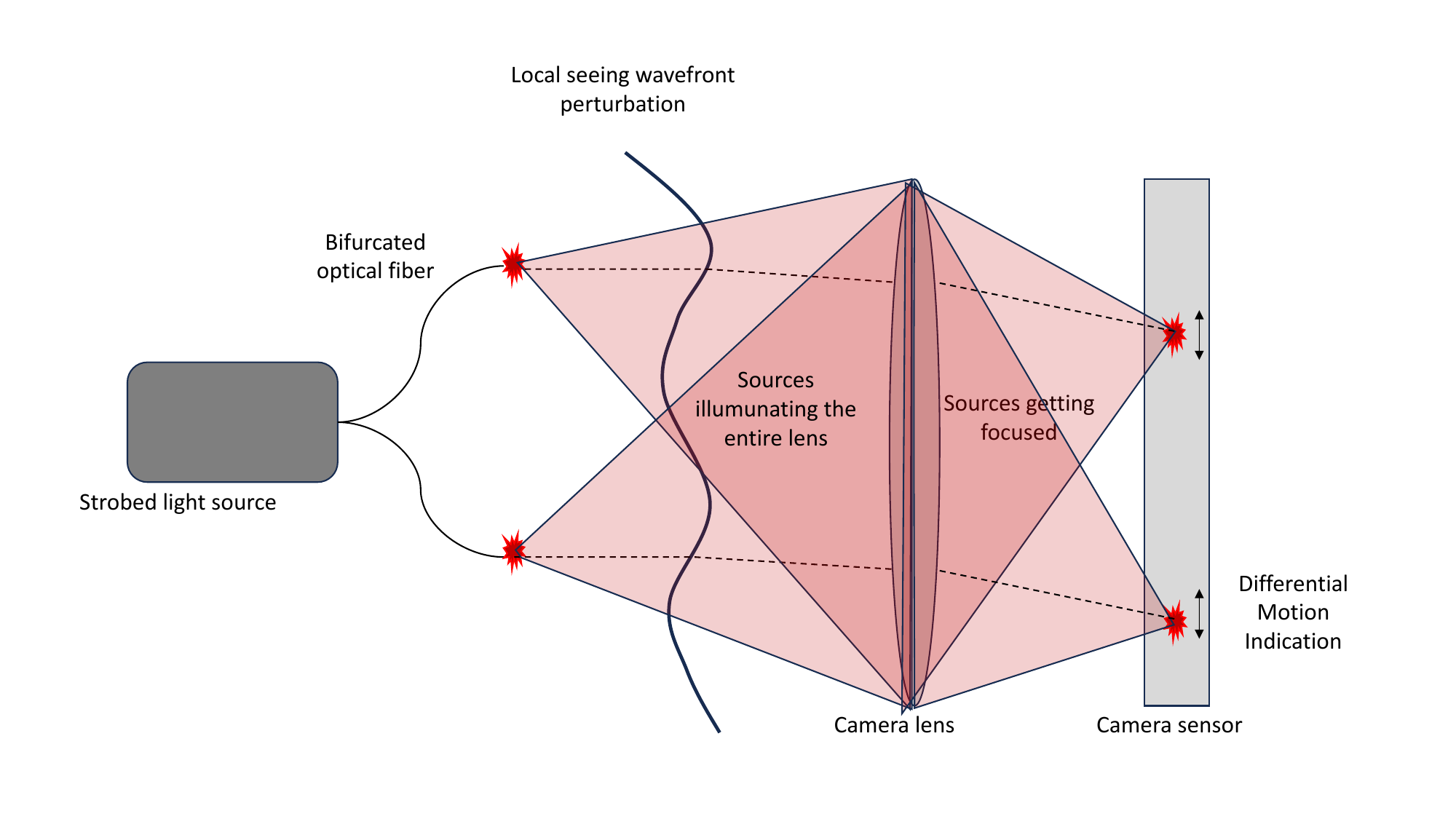}
    \caption{A common pulsed source illuminates two light sources through a bifurcated optical fiber. The light travels through a local wavefront perturbation and proceeds to the lens of a fast shutter speed camera, where it is refocused to the sensor of the camera. The short time scale change of the wavefront perturbation results in changes of angle of arrival of the two light sources between the high speed camera shots. Measuring differential image motion caused by this change probes the difference in perturbation induced tilt due to local (dome) seeing. The short 1/8000 second pulses taken at rate of 7fps in a span of 1.5 seconds let us avoid the issues caused by having a continuous light source in the telescope dome and capture the wavefront in a single configuration.}
    \label{fig:conceptual_dimsum}
\end{figure}

We use a high speed camera to capture the short time scale index of refraction fluctuations that create a local wavefront perturbation. A strobed light source is divided by optical fibers to create multiple sources. The camera takes a picture of each flash which provides a snapshot of the atmosphere at that moment. Rapid change of local wavefront tilt causes each source to reach the camera lens at a different angle which causes a differential motion among the sources. Fig.~\ref{fig:conceptual_dimsum} illustrates the operating principle.

We determine differential image motion by calculating the standard deviation of the distance between the centroids of pairs of sources. The centroids and differential motion of pairs are calculated from a sequence of short-exposure (1/8000 seconds) image bursts that are taken over a span of a few seconds. The differential motion approach isolates the atmospheric effects from common-mode effects such as vibrations in the mount.

Finally, short pulses of light allows the use of this instrument while the telescope is taking data. Flashes at the order of milliseconds at low duty cycle do not end up creating detectable background light in the main instrument. Section~\ref{sec:physical_setup} outlines how we built a physical setup using the principles described.

\subsection{Physical Implementation}\label{sec:physical_setup}
We used a modified Canon 5D Mark IV DSLR camera which can take up to 7 frames per second. The camera had the Bayer filter and IR cutoff filters removed (by MaxMax.com), so it serves as a 30.4 Mpix monochrome imager. The panel containing the sources was 6.5 meters from the camera. We used a Bower 500mm focal length mirror telephoto lens, which was (barely!) able to focus close enough to produce an image of the panel on the focal plane. That image subtends 3.5 degrees. As our strobed light source, we used a 5W Hamamatsu Model L9455-11 Xenon Flash Lamp. We used a Godox X Pro C remote flash trigger (mounted to the camera's flash shoe) and a companion Godox X1R receiver to synchronize the external strobe to the camera shutter opening. A custom circuit was added to the receiver end of the RF shutter controller link. We used a simple voltage divider and bias voltage, connected to the flash receiver, to provide a 10V trigger to the Xenon lamp. 

Unlike the conceptual setup described in Sec.~\ref{sec:conceptual_setup}, we used 16 fiber-tip sources instead of 2. Using 16 sources let us observe the differential image motion with respect to x, y, and total separation. To generate 16 simultaneously strobed sources we used five 1-to-4 200 $\mu$m optical fiber splitters. One of the 1-to-4 fibers was connected to the Hamamatsu lamp making 4 sources. Then, we connected another 1-to-4 optical fiber to each of the sources making a total of 16 sources. We used a pegboard with SMA threads to arrange sources in a way that would give the most efficient data collection. Due to mismatches in light distribution across the optical fibers, there could be differences in the fluxes of the sources. As long as the fluxes are enough to make a high enough SNR, that is not a problem. In fact, a lack of flux has not been a problem with the 5w Hamamatsu Model L9455-11 Xenon Flash Lamp.
We had to insert an OD 1 ND filter between the source and optical fibers using a Thorlabs Multimode Fiber Optic Filter/Attenuator Mount to avoid saturation, even at the lowest-power setting on the lamp. Operating at low light levels also minimizes background light in the dome.

The duration of the light pulses was 125$\mu$sec. To avoid any synchronization problems on site we used an exposure time of 20ms. The effective exposure time is the flash duration, however. The images were taken remotely and automatically using {\tt gphoto2} software \citep{gphoto2}. We took successive bursts of images at 7fps over 24 hours with 5 or 10 minutes between bursts. The ISO of the camera was set to 100, the white balance was 5200K. The images were stored as raw files. The metering mode was evaluative metering mode (the default of the camera). There was no delay between flash and shutter. Depending on the camera and the physical flash lamp, it is possible to use different values for ISO, white balance, and metering. We have found that the ISO=100 gain setting to deliver the best combination of SNR and dynamic range. The lens was focused manually. 


We have installed this device in the Auxiliary Telescope (AuxTel) of the Rubin Observatory \citep{ivezic2019lsst}. The AuxTel is a 1.2m telescope adjacent to the main Rubin telescope, which is serving as a system integration pathbreaker for the Rubin project \citep{ingraham2020vera}. The camera was 2 feet above the ground and 6.5 meters away from the sources. The camera was connected to an Apple MacMini which could be reached remotely. The image taking commands were given to the camera through the MacMini. The burst of images taken with the camera were directly transferred to the MacMini and then moved to an external disk. Finally, they were transferred to the Harvard computing cluster. 

The data analysed in this paper is a combination of two data collection runs. The first run took place from 2023/03/10 19:19 Coordinated Universal Time (UTC) to 2023/03/11 19:19 UTC. The second run was from 2023/03/14 15:38 UTC to 2023/03/16 10:02 UTC. This gave us a total of 3 days and 3 nights of data.

\newpage 
\section{Differential Motion Calculation}\label{sec:analysis}
\subsection{Data Processing From Raw Images To Final Differential Motion Values}\label{subsec:data_processing}
Table \ref{tab:Steps} lists the steps to get differential image motion data from the raw images taken by the setup. The middle column describes the steps in detail and the right column outlines the software used if the type of software is essential to the step.
\begin{longtable}{|c|p{0.65\columnwidth}|p{0.25\columnwidth}|} 
    \hline
    Step & Description & Software Used \\ 
    \hline 
    \multicolumn{3}{|c|}{\textit{Measuring The Positions of The Sources From The Images}} \\
    \hline
    1 & Extract ``time of creation" of the images and store it in a csv file that includes the name of the image file and the exact time of creation. & python {\tt exifread} \citep{exifread} and {\tt pandas} \citep{pandas, pandas_paper} \\
    \hline
    2 & Read the images as a 2D array. & python {\tt rawpy} \citep{rawpy} \\
    \hline
    3 & Subtract the median value of the array from every entry in the array. This is effectively a background subtraction of a very uniform background. & python {\tt numpy} \citep{numpy} \\
    \hline
    4 & Convolve the data with a 2D Gaussian kernel with a standard deviation of 2.5 pixels. We chose a standard deviation of 2.5 pixels because the FWHM of the sources were approximately 5 pixels. 2.5 pixel  ensured the smoothing of the PSF and elimination of ghosting effects.  & python {\tt astropy.convolve} \citep{astropy:2022} \\
    \hline
    5 & Find the centroids of the PSFs of sources using the python version of {\tt source extractor}. After convolvement and background subtraction, a {\tt threshold} of a peak of 20 counts accurately detects sources. To ensure {\tt source extractor} does not read a single source as multiple sources, we set the deblending threshold ({\tt deblend\_nthresh}) to 1.& python {\tt sep.extract} \citep{sex_original, sep} \\
    \hline 
    6 & Compare the number of sources found in a given image with the expected number of sources to filter out bad data. If an image file has the expected number of sources, add the positions of the sources to the data that will be used for further processing. &  \\
    \hline 
    \multicolumn{3}{|c|}{\textit{From Positions of Individual Sources To Differential Motion Between Source Pairs}} \\
    \hline
    7 & Combine the position data from the {\tt source extractor} with file name and file time data. &  {\tt sep.extract} and {\tt pandas}\\
    \hline 
    8 & Calculate the mean position of all the sources in a given image. Represent the position of the sources relative to that mean position. The rest of the calculations are done with respect to the adjusted position. This avoids boresight jitter and other pointing problems. & \\
    \hline 
    9 & Use the time data of the files to separate the image files in terms of bursts. As outlined in Sec.~\ref{sec:physical_setup}, we know the time between successive bursts is at least 5 minutes. Thus, we characterize every image that has been taken in the same minute as a part of the same burst. Considering a burst occurs within 1.5 seconds, making the interval 1 minute lets us safely include every image that was part of that burst. & python {\tt  datetime} and {\tt pandas} \\
    \hline 
    10 & For every image in a burst, take the distance between the centroids of source pairs. & \\ 
    \hline
    11 & Calculate the differential motion between source pairs by taking the standard deviation of distances measured between them within a burst of images. This is the main quantity we measure and is referred to as DIM in Sec~\ref{sec:early_results}.& \\
    \hline
    12 & Calculate the uncertainty of the differential motion. The uncertainty in the position of a source is FWHM divided by the SNR. The SNR is the square root of the flux of the sources. & \\
    \hline
    13 & Select bursts that have at least 7 images to calculate differential motion to get reliable standard deviation values. & \\
    \hline
    14 & Convert the data from units of pixels to angular separations. &  \\
    \hline
    15 & Do additional correction to account for focus affected by global temperature variations in the dome. This step  takes place between step 7 and step 8. It is further explained in Sec.~\ref{sec:correction_of_focus}&  \\
    \hline
    \caption{This table lists the image processing steps used to go from RAW camera images to image motion.}
    \label{tab:Steps}
\end{longtable}

\begin{figure}
    \centering
    \includegraphics[width=0.8\textwidth]{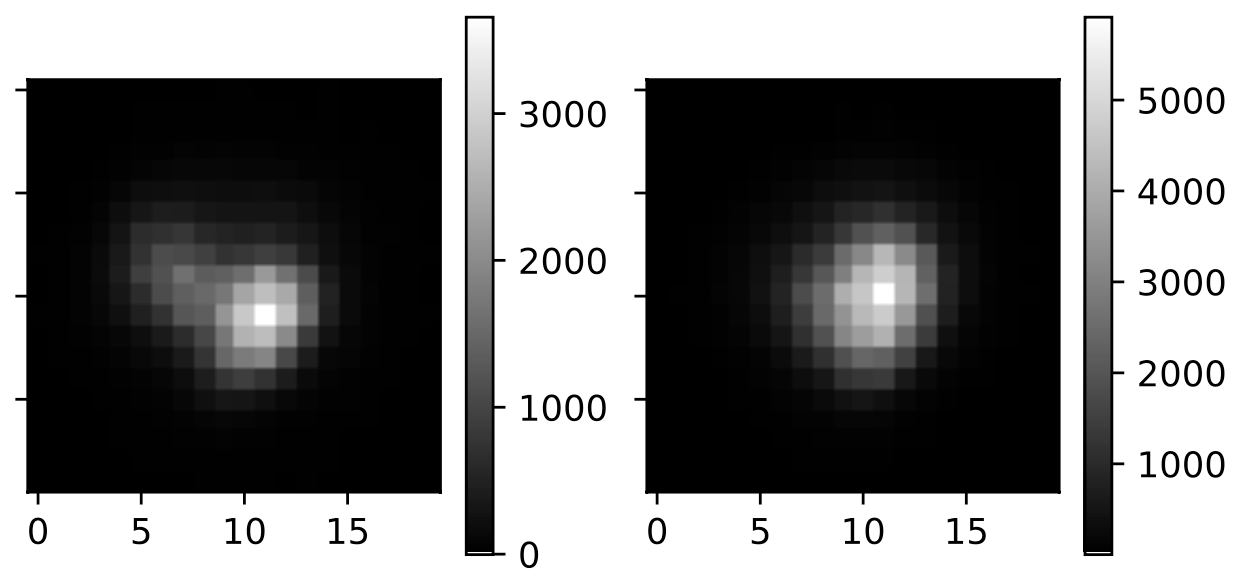}
    \caption{Example PSFs of the sources after background subtraction and before convolution. We present the PSFs in a 20x20 pixel box where the center of the box is the centroid identified by the python version of {\tt source extractor} \citep{sex_original, sep}. The color of each pixel pair represent the amount of flux received. The lighter color represents more flux.}
    \label{fig:2D}
\end{figure}

\begin{figure}
    \centering
    \includegraphics[width=0.8\textwidth]{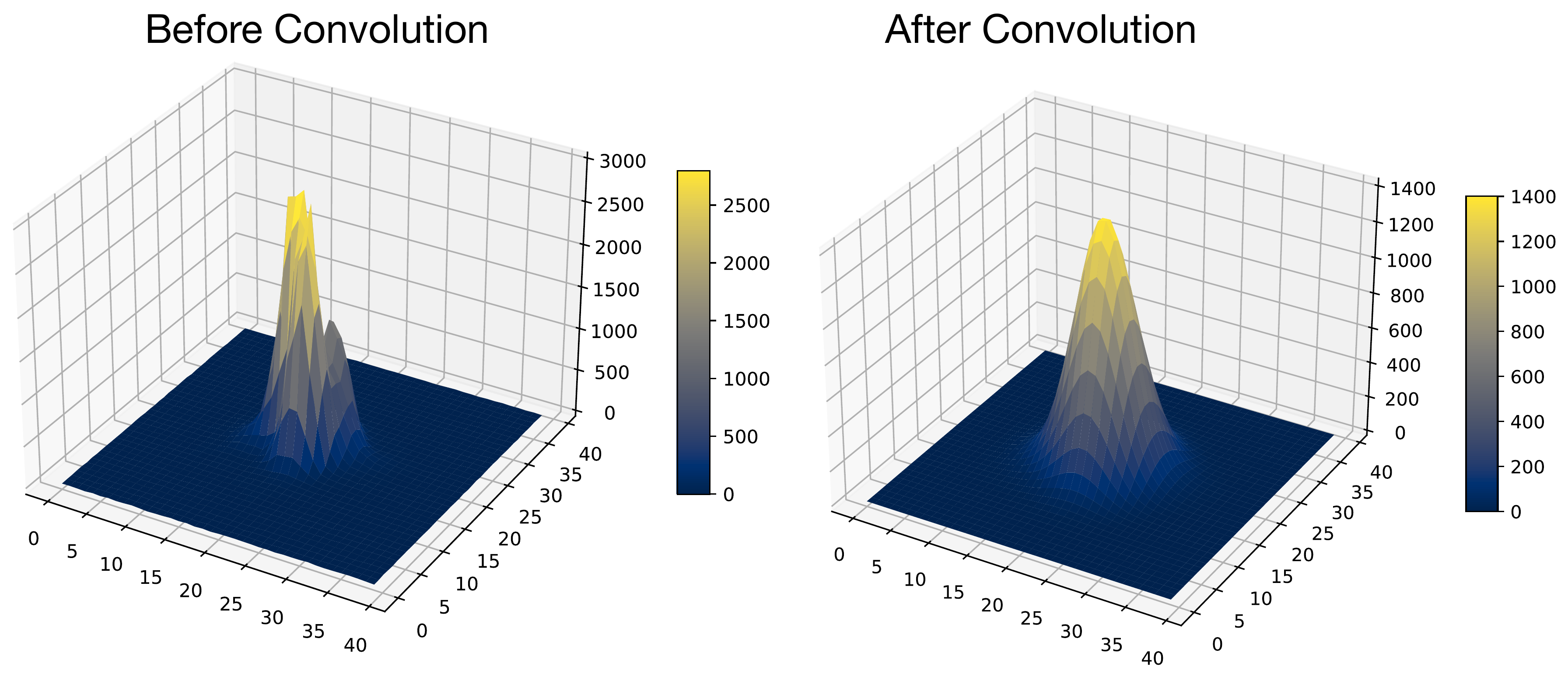}
    \caption{3D plots of the strobed light sources. The horizontal axes are in pixel units. The center of the box is the centroid identified by the python version of {\tt source extractor} \citep{sex_original, sep}. The vertical axis is flux in ADUs. The plot on the left is the PSF of the strobed light source before convolution, with ghosting artifacts. The right plot shows the smoothed PSF after convolution.}
    \label{fig:convolvement}
\end{figure}

Figure~\ref{fig:2D} shows PSFs of two example sources before convolving. Figure~\ref{fig:convolvement} shows an example of how convolving the data smooths out the PSF and eliminates ghosting effects. Given the number of counts, $N$, gain of a few electrons per ADU, and PSF FWHM, a typical centroid uncertainty is of order FWHM/$\sqrt{N_e}\sim10^{-2}$ arcseconds. 

As stated in step 10 of Table~\ref{tab:Steps}, we calculate the differential image motion between a pair of sources by taking the standard deviation of the distances measured between them within a burst of images. We refer to this value as DIM and it makes up the basis of our turbulence measurements. DIM values are ten times larger compared to the centroid uncertainties calculated above. Thus, we conclude that the differential image motion (DIM) we see is dominated by atmospheric effects, not counting statistics. 


\subsection{Reducing Sensitivity to Pointing Errors and Plate Scale Changes}
\label{sec:correction_of_focus}
There are two potential sources of systematic error that we seek to avoid. 
One is pointing error, due to both vibration and slow drift in the camera mount. We eliminate this by computing only differential image motion, so that common mode changes in position cancel out. The second effect we see is a change in plate scale that is driven by thermally induced changes in the effective focal length of the mirror lens. To eliminate this effect, we adopted an image processing scheme that computes differential image motion over short spans of time, namely the few-second long bursts of rapid image collection. Each burst included 10 images on average but it varied from 7 to 15 depending on the state of the camera. With this image processing scheme, the single differential image motion (DIM) values we calculate from bursts that only last few seconds are not affected by plate scale changes. However, effects of plate scale changes become prominent when we do further calculations that include averaging of DIM values over longer time periods. 

We chose to create a scaled coordinate system that compensates for these plate scale changes. Changes in plate scale produce a rubber-sheet distortion of the image on the focal plane. If we center our coordinate system on the mean position of the sources on the image, this manifests as a radially symmetrical distortion of the image.

We computed a scaled, offset coordinate system in order to compensate for these potential sources of systematic error. This was done by:
\begin{enumerate}
\item Compute the center-of-mass of the sources by computing the mean values of x and y for all spots in the image. All subsequent positions are made relative to this center of mass. This avoids boresight jitter and other pointing problems. 
\item Compute an average distance of each individual source from this center of mass, for all the images taken. We denote the mean radial position of the $i$'th source in a given image as $r_i$, and we define the average position of the $i$'th source over the entire run as $\bar{r_i}$.
\item In order to compensate for an overall change in plate scale we then compute a global radial distortion parameter. For every image $n$, we compute the mean ratio of the positions of the sources to the average positions of the sources. Denote this radial distortion parameter as $\Delta r_n$, which is 
\begin{equation}\label{eq:delta_r}
\Delta r_n = \frac{1}{\rm{N}}\sum_{i}^{} |\frac{r_{i_n}}{\bar{r_i}}|
\end{equation} where $r_{i_n}$ is the position of the $i$'th source in the $n$'th image and $N$ is the number of sources.
\item We compensate for this radial distortion by correcting the separations  measured in a given image $n$ with: $r_i \rightarrow \frac{r_i}{\Delta r_n}$. Over a temperature range of 16$^0$C, the maximum correction applied was less than 0.11\%.
\end{enumerate}

\newpage 
\section{Early Results}\label{sec:early_results}

This section presents the early results from DIMSUM based on the two nights of operation described above. After performing boresight offset and plate scale corrections, for each image burst we computed the standard deviation in separation for each of the N(N-1)/2 = 120 image pairs in each frame. That standard deviation is defined as the differential image motion (DIM) for that source pair for that burst, which lasts a few seconds. 

\begin{figure}[h!]
    \centering
    \includegraphics[width=0.7\textwidth]{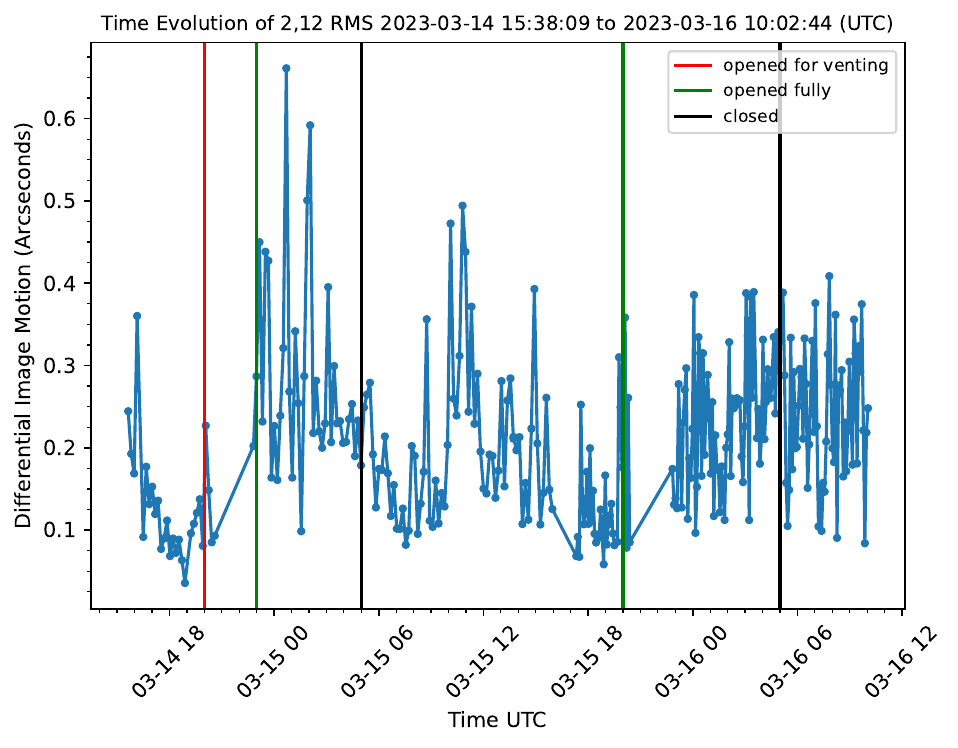}
    \caption{Time evolution of differential image motion (DIM), the standard deviation of the distance between a source pair within a given burst, from 2023/03/14 15:28 to 2023/03/16 07:02. The vertical axis shows the DIM value. The red, green, and black vertical lines respectively indicate the times the dome was opened for venting, fully opened, and closed. We see clear changes in both the DIM measured value and its variance, depending on dome status. The daytime excursion on 3/15 suggests that there may have been an unlogged dome opening. This conjecture will be clarified when we have more data.}
    \label{fig:rms_evolve}
\end{figure}

Figure~\ref{fig:rms_evolve} shows the time evolution of differential motion for a widely separated (3.4 degrees) source pair from 2023/03/14 15:28 to 2023/03/16 10:02 UTC.

A number of conclusions can be drawn from Fig.~\ref{fig:rms_evolve}. The dome status impacts both the mean DIM value and the fluctuations about that mean. Between 21:00 and 23:00 (UTC) on 3/14, the images we obtained do not pass our data quality cuts. This is due to increased background light levels impacting the number of sources we detect. Adaptive image analysis and/or dynamically adjusting the strobed light level are possible ways to contend with this. A similar data gap exists early in the subsequent night. 

\begin{figure}[h!]
    \centering
    \includegraphics[width=\textwidth]{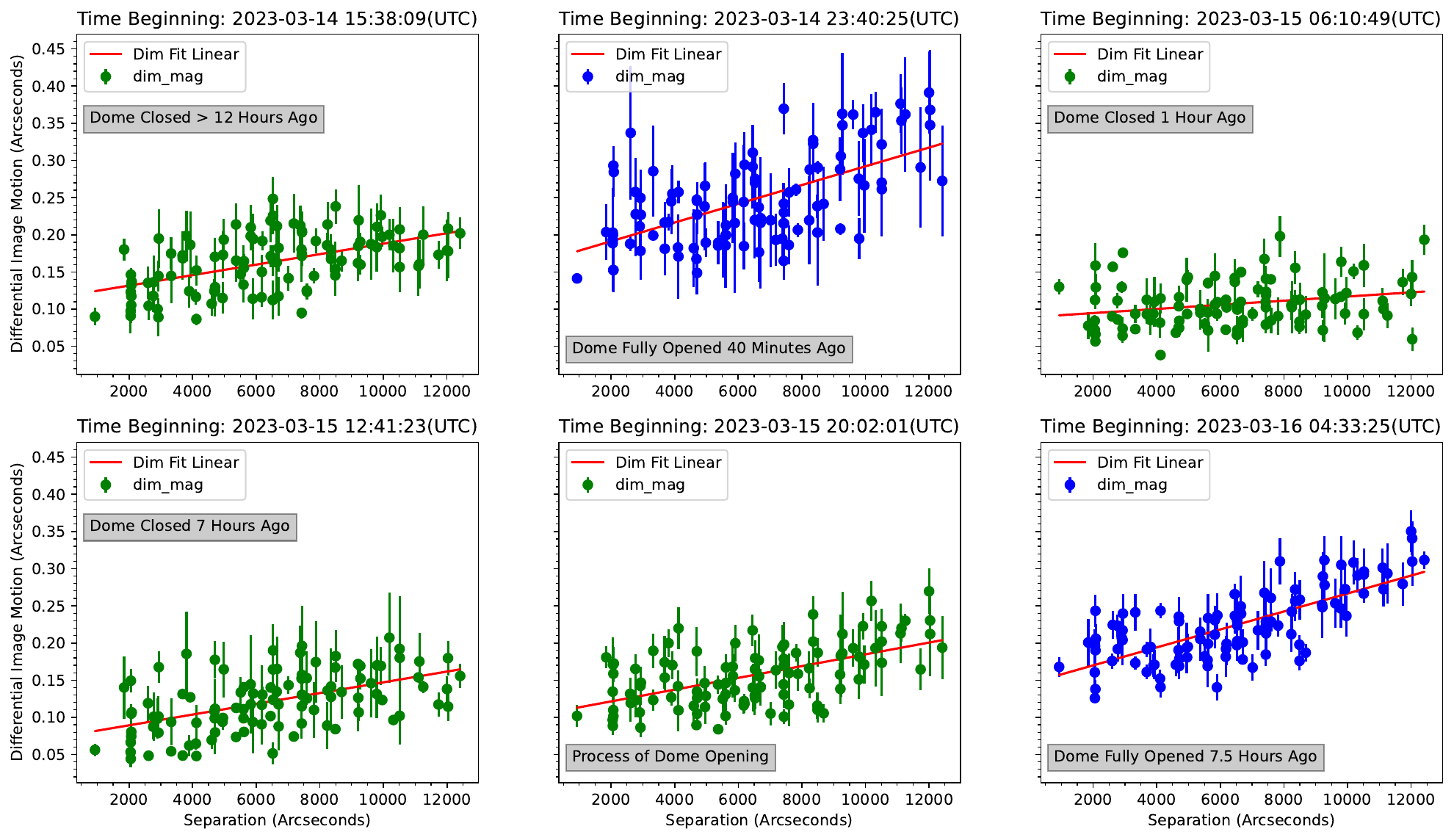}
    \caption{Examples of DIM$_{30}$ values, per source pair, plotted against the separation between source pairs, for different half hour intervals. Every data point represents a single source pair for that interval. The horizontal axes are the pair separations in pixels. The vertical axis shows DIM$_{30}$ values in RMS arcseconds, with error bars that reflect the standard deviation of the points that were averaged to obtain DIM$_{30}$. The red line is the linear fit to the relation of DIM$_{30}$  and separation. The titles of each panel indicate the beginning time of the 30 minute averaging interval. The panels with green and blue data points respectively represent intervals when the dome was closed and open. The gray boxes provide further information about the state of dome during the interval shown in the panels. We see that both the y-intercepts and slopes of the data are greater when the dome is open.}
    \label{fig:dim_graphs}
\end{figure}

Since we are looking for long-term trends, we partitioned the DIMSUM data into 30 minute intervals, and computed a 30-minute average DIM$_{30}$ per source pair for each time block. The scatter in those measurements was used to assign an uncertainty to the DIM$_{30}$ data points. We took advantage of DIMSUM's multi-source configuration to explore the dependence of DIM on source separation. Figure~\ref{fig:dim_graphs} shows plots of how DIM$_{30}$ values depend on the separation (in arcseconds) between source pairs, for different 30 minute intervals. 

Figure~\ref{fig:dim_graphs} clearly shows a linear relationship between separation and differential image motion for source pairs.  In order to extract a parameter that captures the characteristic level of index variation we performed a linear fit for the DIM$_{30}$ vs. separation data, for each interval. We chose to use the y-intercept and DIM$_{30}$ value at 3 degrees (or 5000 pixels) of angular separation to characterize the index variations. The panel of figures suggests a measurable difference in these parameters when the dome is open vs. closed. 

We had previously found in the lab that the DIM values reached a plateau at large separations, when the index variations along the two sight lines become uncorrelated. We see less of this behaviour when DIMSUM is operated inside the AuxTel dome. The spatial correlation scale for index variations appears to be larger in the telescope dome than it was in the laboratory. The dome-closed data do hint at a rollover at the largest separations, but the system would clearly benefit from more standoff distance and/or wider spatial separation between sources. We think it would be instructive to be able to measure that correlation scale, and will strive to do so in the next generation of DIMSUM implementation. 





\begin{figure}[h!]
    \centering
    \includegraphics[width=0.7\textwidth]{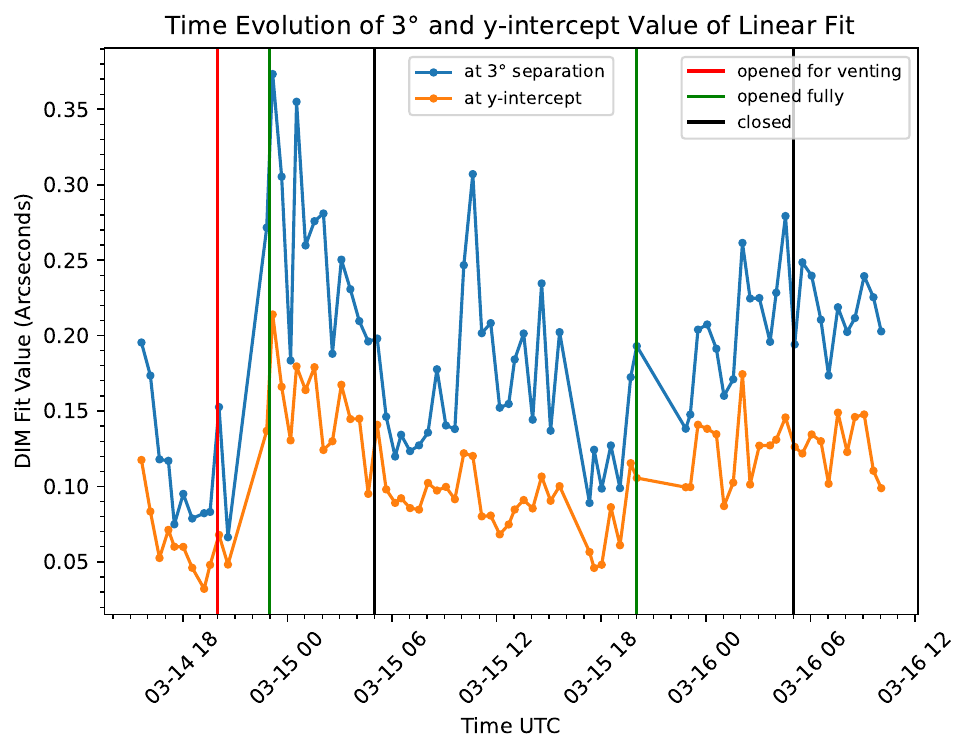}
    \caption{Time evolution of the fitted y-intercept and the averaged DIM value at 3$^{\circ}$ separation of the linear fits to the source separation, over 30 minute intervals. The vertical axis shows the DIM fit values in RMS arcseconds. The blue and orange lines respectively represent fit values at a distance of 3$^{\circ}$ and 0$^{\circ}$ (y-intercept of the fit). The red, green, and black vertical lines respectively indicate the times the dome was opened for venting, fully opened, and closed. These two proxies for dome seeing are well correlated, and while there are confounding factors (again that mid-day interval when we suspect the dome was opened) there is a correlation with dome state. The amount of differential image motion we measure, roughly 0.25 arcsec, is a lower bound on the dome seeing contribution to image FWHM since we only measure angle-of-arrival variations over a relatively small range of spatial wavenumber $k$.}
    \label{fig:value_evolve}
\end{figure}

Figure~\ref{fig:value_evolve} shows the time evolution of both the y-intercept and the fitted DIM$_{30}$ value at a source separation of 3 degrees (5000 pixels), for each 30 minute interval in the data set.

Because the AuxTel science instrument images are dominated by astigmatism and were obtained in a wide variety of modes (imaging in different passbands, spectroscopy, etc) we can't meaningfully correlate the DIMSUM dome seeing proxies with image FWHM. That will have to wait for a more consistent and longer time-span data set. We can however perform a comparison of the measured DIMSUM dome seeing parameters and data obtained from an acoustic anemometer that was operating within the dome at the same time. 

The acoustic anemometer not only provides a measurement of the local wind velocity vector, it also uses the mean sound speed to produce an estimate of ``acoustic temperature''. Because this measurement is obtained with ultrasonic travel times on a small (10 cm) parcel of air, the thermal time constants involved are very short. The Campbell Scientific CSAT3B-NC 3D acoustic anemometer takes data at 10 Hz. The standard deviation of ten successive acoustic temperature measurements is a direct measurement of variations in temperature, which in turn drive variations in the optical index of refraction of that air parcel. We therefore consider the standard deviation in acoustic temperature as another indicator of the amount of dome seeing, albeit at a single location rather than the weighted line integral measured by DIMSUM. 

\begin{figure}[h!]
    \centering
    \includegraphics[]{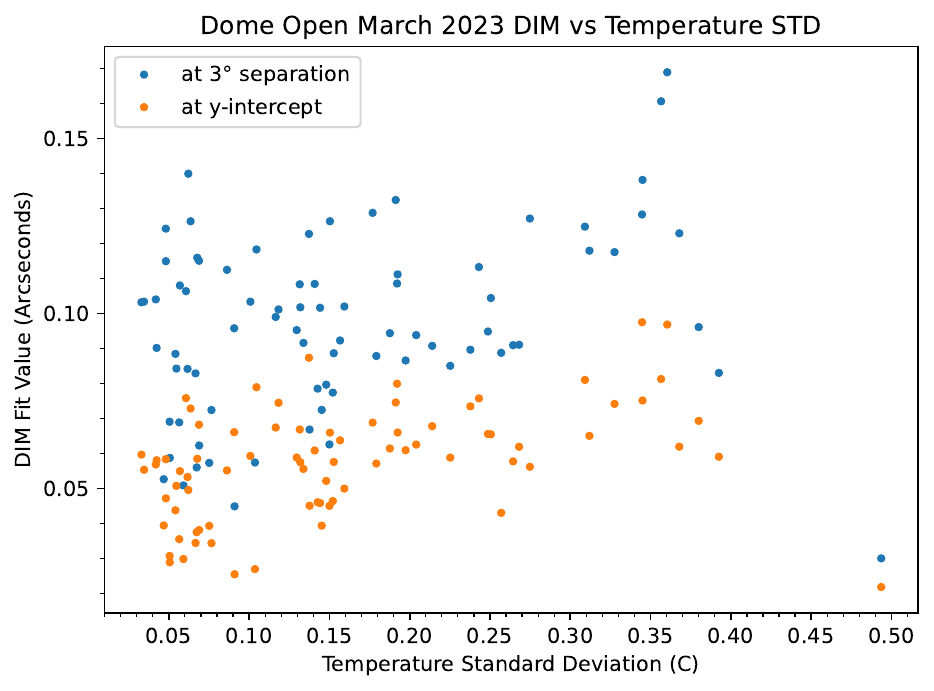}
    \caption{Correlation of DIMSUM dome seeing parameters (y-intercept and the value at 3$^{\circ}$ separation) with the standard deviation of acoustic temperature. The horizontal axis indicates the standard deviation of the sonic temperature measured in degrees Celsius. The vertical axis indicates the two DIMSUM dome seeing parameter values in arcseconds. The blue and orange points respectively represents linear fit values at y-intercept and a distance of 3$^{\circ}$. We see a clear  positive correlation between these two dome seeing indicators. The outlier point in the lower right remains enigmatic.}
    \label{fig:temp_std}
\end{figure}

Figure~\ref{fig:temp_std} shows the comparison of the DIMSUM dome seeing parameters with the standard deviation of the acoustic temperature, from the DIMSUM data collected when the dome was open through 2023/03/10, 2023/03/14, 2023/03/15, 2023/03/16. Figure~\ref{fig:temp_std} suggests a positive correlation between sonic temperature standard deviation and DIMSUM dome seeing parameters. Both the fitted values at 3 degree separation and the y intercepts increase with increasing acoustic temperature standard deviation. This positive correlation verifies, at least on qualitative level, we are measuring the desired quantity with our setup. 

\newpage 
\section{Discussion and Next Steps}\label{sec:discussion}

Our principal objective was to build a measurement system that can provide a quantity that has consistent temporal and spatial correlation with dome seeing, over a substantial optical path length in the dome. Such a measurement, without the need of calibration, will allow for iterative correction of dome seeing by engineering the airflow in the dome. Making airflow and operational adjustments that minimize this measured quantity should drive down the dome seeing contribution to the image quality budget, as long as we have adequate sensitivity. Our preliminary results indicate that we have met this primary goal. The dome seeing parameters that DIMSUM provides do correlate with the contemporaneous standard deviation of acoustic temperature, establishing a clear physical link that is consistent with our expectations. 

We can potentially extract more information from the DIMSUM data than has been explored here. Adding scintillation measurements (recall all the strobed sources have a common flash lamp driving them) would allow a MASS-DIMM like analysis. Increasing the separation between the sources would allow us to exploit the plateau regime and characterize the correlation length scale of index variations. Other changes we envision include: 

\begin{itemize}
    \item Adding an autofocus or remote focus function to the lens. We have found that the system does go out of focus in the face of large temperature changes. 
    \item Perhaps adding a narrowband filter to select the NIR peaks in the Xenon spectrum, and filtering out blue light from the twilight sky. 
    \item Running a complete suite of diagnostic tools, including an external DIMM, to more fully separate upper atmospheric from dome seeing.
    \item Increasing the path length over which DIMSUM operates. This is essentially free gain. This will have to await installation in the much larger main Rubin telescope enclosure.
    \item Correlating the FWHM of images taken by the Auxtel with measurements from DIMSUM to directly link our work to observational outcomes.
\end{itemize}

In conclusion, despite its current limitations, the DIMSUM setup has not only met its initial goals but also set the stage for further possibilities of improving image quality. We envision DIMSUM playing a central role in helping us learn how to operate the Rubin survey system in a fashion that minimizes dome seeing. We can experiment with tailoring the airflow through the enclosure, with optimizing air conditioning set-points, and with choosing the azimuth at which we take data compared to the prevailing wind conditions. Having the ability to factor our other time and condition-dependent contributions to the image quality budget by making an explicit measurement of dome seeing should speed the process of learning how to best operate the Rubin system. 

\section{Data Availability}
The data presented in the figures are available upon request from the corresponding author.
\section{Acknowledgements}
We gratefully acknowledge support from the U.S. Department of Energy under Cosmic Frontier award DE-SC0007881, and from Harvard University. GPT-4 \citep{openai_gpt4} was used to refine this manuscript. We thank Sasha Brownsberger, Jim MacArthur, and Eske Pedersen for helpful comments and suggestions and Craig Lage, Bruno Quint, Tiago Ribiero, and Karla Aubel for their assistance at the Rubin site in Chile.

\bibliography{references}

\begin{thebibliography}{}
\expandafter\ifx\csname natexlab\endcsname\relax\def\natexlab#1{#1}\fi
\providecommand{\url}[1]{\href{#1}{#1}}
\providecommand{\dodoi}[1]{doi:~\href{http://doi.org/#1}{\nolinkurl{#1}}}
\providecommand{\doeprint}[1]{\href{http://ascl.net/#1}{\nolinkurl{http://ascl.net/#1}}}
\providecommand{\doarXiv}[1]{\href{https://arxiv.org/abs/#1}{\nolinkurl{https://arxiv.org/abs/#1}}}

\bibitem[{{Astropy Collaboration} {et~al.}(2022){Astropy Collaboration}, {Price-Whelan}, {Lim}, {Earl}, {Starkman}, {Bradley}, {Shupe}, {Patil}, {Corrales}, {Brasseur}, {N{"o}the}, {Donath}, {Tollerud}, {Morris}, {Ginsburg}, {Vaher}, {Weaver}, {Tocknell}, {Jamieson}, {van Kerkwijk}, {Robitaille}, {Merry}, {Bachetti}, {G{"u}nther}, {Aldcroft}, {Alvarado-Montes}, {Archibald}, {B{'o}di}, {Bapat}, {Barentsen}, {Baz{'a}n}, {Biswas}, {Boquien}, {Burke}, {Cara}, {Cara}, {Conroy}, {Conseil}, {Craig}, {Cross}, {Cruz}, {D'Eugenio}, {Dencheva}, {Devillepoix}, {Dietrich}, {Eigenbrot}, {Erben}, {Ferreira}, {Foreman-Mackey}, {Fox}, {Freij}, {Garg}, {Geda}, {Glattly}, {Gondhalekar}, {Gordon}, {Grant}, {Greenfield}, {Groener}, {Guest}, {Gurovich}, {Handberg}, {Hart}, {Hatfield-Dodds}, {Homeier}, {Hosseinzadeh}, {Jenness}, {Jones}, {Joseph}, {Kalmbach}, {Karamehmetoglu}, {Ka{l}uszy{'n}ski}, {Kelley}, {Kern}, {Kerzendorf}, {Koch}, {Kulumani}, {Lee}, {Ly}, {Ma}, {MacBride}, {Maljaars}, {Muna}, {Murphy}, {Norman}, {O'Steen},
  {Oman}, {Pacifici}, {Pascual}, {Pascual-Granado}, {Patil}, {Perren}, {Pickering}, {Rastogi}, {Roulston}, {Ryan}, {Rykoff}, {Sabater}, {Sakurikar}, {Salgado}, {Sanghi}, {Saunders}, {Savchenko}, {Schwardt}, {Seifert-Eckert}, {Shih}, {Jain}, {Shukla}, {Sick}, {Simpson}, {Singanamalla}, {Singer}, {Singhal}, {Sinha}, {Sip{H{o}}cz}, {Spitler}, {Stansby}, {Streicher}, {{{S}}umak}, {Swinbank}, {Taranu}, {Tewary}, {Tremblay}, {Val-Borro}, {Van Kooten}, {Vasovi{'c}}, {Verma}, {de Miranda Cardoso}, {Williams}, {Wilson}, {Winkel}, {Wood-Vasey}, {Xue}, {Yoachim}, {Zhang}, {Zonca}, \& {Astropy Project Contributors}}]{astropy:2022}
{Astropy Collaboration}, {Price-Whelan}, A.~M., {Lim}, P.~L., {et~al.} 2022, apj, 935, 167, \dodoi{10.3847/1538-4357/ac7c74}

\bibitem[{Barbary(2016)}]{sep}
Barbary, K. 2016, Journal of Open Source Software, 1, 58, \dodoi{10.21105/joss.00058}

\bibitem[{{Bertin} \& {Arnouts}(1996)}]{sex_original}
{Bertin}, E., \& {Arnouts}, S. 1996, \aaps, 117, 393, \dodoi{10.1051/aas:1996164}

\bibitem[{Bustos \& Tokovinin(2018)}]{Bustos_Tokovinin_2018}
Bustos, E., \& Tokovinin, A. 2018, in Ground-based and Airborne Telescopes VII, Vol. 10700, SPIE, 231--240, \dodoi{10.1117/12.2309652}

\bibitem[{{Gilda} {et~al.}(2022){Gilda}, {Draper}, {Fabbro}, {Mahoney}, {Prunet}, {Withington}, {Wilson}, {Ting}, \& {Sheinis}}]{Gilda_2022}
{Gilda}, S., {Draper}, S.~C., {Fabbro}, S., {et~al.} 2022, \mnras, 510, 870, \dodoi{10.1093/mnras/stab3243}

\bibitem[{{gphoto2}(2022)}]{gphoto2}
{gphoto2}. 2022, {The gphoto2 commandline frontend}.
\newblock \url{http://www.gphoto.org/proj/gphoto2/}

\bibitem[{{Hardy}(1998)}]{Hardy_1998}
{Hardy}, J.~W. 1998, {Adaptive Optics for Astronomical Telescopes}

\bibitem[{Harris {et~al.}(2020)Harris, Millman, van~der Walt, Gommers, Virtanen, Cournapeau, Wieser, Taylor, Berg, Smith, Kern, Picus, Hoyer, van Kerkwijk, Brett, Haldane, del R{\'{i}}o, Wiebe, Peterson, G{\'{e}}rard-Marchant, Sheppard, Reddy, Weckesser, Abbasi, Gohlke, \& Oliphant}]{numpy}
Harris, C.~R., Millman, K.~J., van~der Walt, S.~J., {et~al.} 2020, Nature, 585, 357, \dodoi{10.1038/s41586-020-2649-2}

\bibitem[{{Hickson}(2014)}]{Hickson_2014}
{Hickson}, P. 2014, \aapr, 22, 76, \dodoi{10.1007/s00159-014-0076-9}

\bibitem[{Ingraham {et~al.}(2020)Ingraham, Clements, Ribeiro, Reuter, Fisher-Levine, Hoblitt, Lupton, Thomas, Stubbs, Arndt, {et~al.}}]{ingraham2020vera}
Ingraham, P., Clements, A.~W., Ribeiro, T., {et~al.} 2020, in Software and Cyberinfrastructure for Astronomy VI, Vol. 11452, SPIE, 124--139

\bibitem[{Ivezi{\'c} {et~al.}(2019)Ivezi{\'c}, Kahn, Tyson, Abel, Acosta, Allsman, Alonso, AlSayyad, Anderson, Andrew, {et~al.}}]{ivezic2019lsst}
Ivezi{\'c}, {\v{Z}}., Kahn, S.~M., Tyson, J.~A., {et~al.} 2019, The Astrophysical Journal, 873, 111

\bibitem[{{Lai} {et~al.}(2019){Lai}, {Withington}, {Laugier}, \& {Chun}}]{Lai_2019}
{Lai}, O., {Withington}, K., {Laugier}, R., \& {Chun}, M. 2019, arXiv e-prints, arXiv:1912.02680, \dodoi{10.48550/arXiv.1912.02680}

\bibitem[{OpenAI(2023)}]{openai_gpt4}
OpenAI. 2023, GPT-4: Generative Pre-trained Transformer 4

\bibitem[{pandas developers(2023)}]{pandas}
pandas developers. 2023, pandas-dev/pandas: Pandas, latest,  Zenodo, \dodoi{10.5281/zenodo.3509134}

\bibitem[{Racine {et~al.}(1991)Racine, Salmon, Cowley, \& Sovka}]{Racine_1991}
Racine, R., Salmon, D., Cowley, D., \& Sovka, J. 1991, Publications of the Astronomical Society of the Pacific, 103, 1020, \dodoi{10.1086/132920}

\bibitem[{Riechert(2022)}]{rawpy}
Riechert, M. 2022, {rawpy: A Python wrapper for the LibRaw library}, 0.17.2.
\newblock \url{https://pypi.org/project/rawpy/}

\bibitem[{{Stubbs}(2021)}]{Stubbs_2021}
{Stubbs}, C.~W. 2021, \mnras, 508, 3936, \dodoi{10.1093/mnras/stab2781}

\bibitem[{Sévi(2020)}]{exifread}
Sévi, I. 2020, {ExifRead: Read Exif metadata from TIFF and JPEG files}, 2.3.2.
\newblock \url{https://pypi.org/project/ExifRead/}

\bibitem[{{W}es {M}c{K}inney(2010)}]{pandas_paper}
{W}es {M}c{K}inney. 2010, in {P}roceedings of the 9th {P}ython in {S}cience {C}onference, ed. {S}t\'efan van~der {W}alt \& {J}arrod {M}illman, 56 -- 61, \dodoi{10.25080/Majora-92bf1922-00a}

\bibitem[{{Woolf}(1979)}]{Woolf_1979}
{Woolf}, N. 1979, \pasp, 91, 523, \dodoi{10.1086/130532}

\bibitem[{{Zago}(1997)}]{Zago_1997}
{Zago}, L. 1997, in Society of Photo-Optical Instrumentation Engineers (SPIE) Conference Series, Vol. 2871, Optical Telescopes of Today and Tomorrow, ed. A.~L. {Ardeberg}, 726--736, \dodoi{10.1117/12.269103}

\end{thebibliography}
\bibliographystyle{aasjournal}



\end{document}